\documentclass{article}
\usepackage{LaThuileFPSpro}
\usepackage{graphicx}
\usepackage{amsmath}

\def\EPJ                     {\mbox{Eur. Phys. J.}}

\def\JHEP                    {\mbox{JHEP}}
\def\JPhys                   {\mbox{Journal of Phys.}}
\def\NIM                     {\mbox{Nucl. Instrum. and Methods}}
\def\NuclPhys                {\mbox{Nucl. Phys.}}
\def\NuclPhysPS              {\mbox{Nucl. Phys. Proc. Suppl.}}

\def\PhysLett                {\mbox{Phys. Lett.}}

\def\PhysRev                 {\mbox{Phys. Rev.}}
\def\PRL                     {\mbox{Phys. Rev. Lett.}}

\newcommand{\Journal}[4]     {{#1} {\bf {#2}}, {#3} ({#4})}

\newcommand{\arXiv}[1]       {{\tt arXiv:{#1}}}
\newcommand{\WWWAddr}[1]     {{\tt {#1}}}
\newcommand{\FermiPub}[2]    {Fermilab-Pub-{#1}/{#2}}

\newcommand{\mygraphics}[1]{%
  \includegraphics[width=\linewidth,clip=false]{#1}\vspace{0in}
}

\newcommand{\etal}      {\mbox{\it et al.}}

\newcommand {\Dzero}    {\mbox{D0}}

\newcommand {\unitexp}[2] {\mbox{{#1}$^{\mathrm{#2}}$}}
\newcommand {\scinot}[2]  {\mbox{{#1}$\times$10$^{\mathrm{#2}}$}}
\newcommand {\BR}         {\mbox{$\mathcal{B}$}}
\newcommand {\ra}         {\mbox{$\rightarrow$}}

\newcommand {\fbinv}      {\mbox{\unitexp{fb}{-1}}}
\newcommand {\instLunit}  {\unitexp{cm}{-2}\unitexp{s}{-1}}

\newcommand {\pt}         {\mbox{$p_T$}}

\newcommand {\CP}         {\mbox{$CP$}}
\newcommand {\dms}        {\mbox{$\Delta m_s$}}
\newcommand {\dGs}        {\mbox{$\Delta \Gamma_s$}}
\newcommand {\dGcp}       {\mbox{$\Delta \Gamma_{CP}$}}
\newcommand {\phis}       {\mbox{$\phi _s$}}
\newcommand {\effDsq}     {\mbox{$\varepsilon D^2$}}
\newcommand {\taufs}      {\mbox{$\tau_{fs}$}}
\newcommand {\taueven}    {\mbox{$\tau _{even}$}}
\newcommand {\BReven}     {\mbox{$\BR _{even}$}}
\newcommand {\Aslmu}      {\mbox{$A_{SL}^{\mu\mu}$}}
\newcommand {\Asls}       {\mbox{$A_{SL}^{s}$}}

\newcommand {\dmd}        {\mbox{$\Delta m_d$}}

\newcommand {\Vckm}[1]     {\mbox{$V_{#1}$}}

\newcommand {\ppb}      {\mbox{$p\bar{p}$}}
\newcommand {\epem}     {\mbox{$e^+ e^-$}}

\newcommand {\Hadron}[2]     {\mbox{${#1}_{#2}$}}
\newcommand {\Hadronbar}[2]  {\mbox{${\bar{#1}}_{#2}$}}

\newcommand {\Bs}         {\Hadron{B}{s}}
\newcommand {\Bsbar}      {\Hadronbar{B}{s}}

\newcommand {\Ds}         {\Hadron{D}{s}}
\newcommand {\Dsm}        {$D_s^-$}
\newcommand {\Dsp}        {$D_s^+$}

\newcommand {\BsJpsiPhi}  {\mbox{\Bs \ra $J/\psi \phi$}}
\newcommand {\Bsmumu}     {\mbox{\Bs \ra $\mu^+ \mu^-$}}
\newcommand {\Bsmumuphi}  {\mbox{\Bs \ra $\mu^+ \mu^- \phi$}}
\newcommand {\BsDsDs}     {\mbox{\Bs \ra $D_s^{(*)} D_s^{(*)}$}}

\begin{document}

\title{ 
  $B_s$ PHYSICS AT CDF AND D\O
  }
\author{
  Harold G. Evans\\
  {\em Physics Department, Indiana University, Bloomington, IN, 47405, USA}\\
  (for the CDF and \Dzero\ Collaborations)
  }
\maketitle

\baselineskip=11.6pt

\begin{abstract}
Run II at the Tevatron has seen an explosion of results related to the
\Bs\ meson, ranging from tests of QCD models, to probes of
electro-weak symmetry breaking, to direct searches for new physics
effects. I will briefly summarize the CDF and \Dzero\ \Bs -physics
programs, 
describing the suitability of the detectors for doing this kind of
physics, and pointing out how our
knowledge of important quantities has improved through Run II
measurements. 
\end{abstract}

\newpage
\baselineskip=14pt

\section{Introduction}
Who would have thought it would be so fruitful?\footnote{
  In fact, many people have recognized the utility of the \Bs\ meson
  for years now, but it did come as some surprise to the CDF and
  \Dzero\ B-Physics communities that \Bs\ analyses have proven to be
  such a dominant component of the Tevatron B-Physics program.
}
In the past two years, CDF and \Dzero\ have produced over 35 separate
results using \Bs\ mesons. This represents the largest component of the
Tevatron B-physics program, and, indeed, puts the \Bs\ meson in the
position  of being the second-most prolific particle, in terms of physics
output, in the Tevatron zoo -- right after the top quark. This handy
hadron has been used to study such diverse topics as QCD model
building, physics beyond the Standard Model, the electro-weak symmetry
breaking mechanism, and \CP\ violation, to name just a few. In all of
these areas, recent \Bs\ results have sharpened our knowledge of the
Standard Model and its weaknesses substantially.

Success comes at a price though -- at least for conference
audiences. The overwhelming number of results means that we cannot
discuss any of their beautiful facets in detail; 
this article is already long enough.
We will therefore concentrate on emphasizing the breadth of
physics issues addressed by studies of the \Bs\ meson, and will show
how our understanding of these issues has improved since data from
Run II at the Tevatron has been analyzed.

Obviously, this represents a snapshot of the Tevatron \Bs -physics
program, taken with up to 1.3 \fbinv\ of data. The Fermilab
Tevatron accelerator continues to deliver proton-antiproton collisions
to CDF and \Dzero\ at a furious pace. At the time of the conference
each experiment had recorded over 2 \fbinv\ of data, with more rolling
in every day. The final section of these proceedings will then be
devoted to a brief discussion of what we plan to do with the bounty of
\Bs\ mesons that will come with this additional data.

\section{The \Bs\ at CDF and \Dzero }
Before we embark upon a description of results, it's instructive to
examine the capabilities of the Tevatron collider experiments in areas
important to the study of \Bs\ mesons. 
Both CDF\cite{cdfrun2} and \Dzero \cite{d0run2} are well
suited to take advantage of the large number of \Bs\ mesons produced
in proton-antiproton collisions at the Tevatron. The \Bs\ production
rate within the CDF and \Dzero\ detector acceptance is around 600 Hz,
at luminosities of \scinot{2}{32} \instLunit .
This can be
compared to around 1 Hz at the B-factories, when they run at the
$\Upsilon$(5S) resonance;
and 5000 Hz at
the upcoming LHCb experiment. 
Thus the Tevatron is the only facility where \Bs\ mesons are produced
currently in any large numbers.

This large \Bs\ production rate, however, is dwarfed by
the rate of \ppb\ bunch crossings (1.7 MHz)
meaning that triggers are critical to the B-physics program at the
Tevatron. 
Both CDF and \Dzero\ use three-level trigger systems and rely heavily
(almost exclusively in \Dzero 's case) on single or di-lepton signals
to collect samples of \Bs\ mesons. \Dzero\ makes use of its excellent
muon detectors to construct low \pt\ threshold single- and di-muon
triggers without resorting to muon impact parameter cuts,
which bias decay length distributions,
except at the highest Tevatron luminosities.
CDF uses both electrons and muons for B-physics triggers, but
applies impact parameter cuts to most single-lepton triggers. 

Because the CDF trigger system is capable of accepting events from its
first level at a rate of up to 30 kHz
(the corresponding \Dzero\ level-1 bandwidth limit is around 2 kHz),
the CDF collaboration has been able to design triggers that select,
at level-1, events with two tracks forming a vertex
displaced from the primary \ppb\ interaction point.
Such events are enriched with fully hadronic decays of \Hadron{B}{}
mesons, which allows CDF access to the wide range of important studies
that can be done using these decay modes.

For many of the analyses discussed in this article the
primary means of identifying \Bs\ mesons is through their semileptonic
decay: 
\Bs \ra \Dsm $\ell^+ \nu_{\ell} X$.
This decay has a branching fraction of 7.9\%\cite{pdg06}, and its
identification highlights many of the experimental challenges that
face CDF and \Dzero .
To begin with, there is the issue of lepton identification. \Dzero\ has
the advantage here because of their muon detectors, which extend to
pseudo-rapidities ($\eta = -\ln [ \tan (\theta / 2) ]$) between
$\pm$2, while CDF's muon system covers only 
$|\eta | < 1$. 
Additionally, the \Dzero\ muon system is shielded by 12--18
interaction lengths of material
(a factor of around three more than CDF's)
and includes toroidal magnets for local muon momentum measurements.
Taken together, these reduce many muon background sources in \Dzero\ 
to a negligible level.

Identification of \Ds\ meson decays, on the other hand, requires
excellent tracking capabilities. The experiments use 
$\phi (K^+ K^- ) \pi^+$,
$\bar{K}^{*0} (K^- \pi^+ ) K^+$,
$K^0_S (\pi^+ \pi^- ) K^+$, and
$\pi^+ \pi^- \pi^+$
decay modes of the \Dsp\ (and charge conjugates) in a range of
analyses. CDF has the edge here, 
mainly because of the larger volume of their tracking system 
(extending from radii of 1.5 -- 137 cm, as compared to the \Dzero\
tracker, which spans 2.8 -- 52 cm)
and the
larger number of space-point measurements it makes
(normally more than 100 for CDF, but only $\sim$20 for \Dzero ). 
This allows CDF to reconstruct multi-particle invariant masses with
significantly better accuracy than \Dzero , making combinatorial
backgrounds less of a problem.

Finally, both detectors reconstruct displaced vertices with similar
resolution, although CDF does a slightly better job because their
tracking extends to lower radii\footnote{
  Note that \Dzero\ added a {\it Layer-0} silicon detector at a
  radius of 1.7 cm in June, 2006. Results using this new detector were
  not yet available for this conference though.
}.
As we will see, resolution is particularly important in \Bs -mixing
analyses, where time structures on the order of 100 fs must be
reconstructed. Both the CDF and \Dzero\ tracking systems are capable
of this feat in the reconstruction of semileptonic \Bs\ decays, having
average resolutions in the 15-20 fs range. However, because of CDF's
large fully hadronic \Bs\ decay sample, they are also able to take
advantage of the better vertexing resolution ($<$10 fs) achievable in
these types of decay.

Using these upgraded detectors,
both experiments have been able to accumulate 
\Bs\ meson data samples
that contain orders of magnitude more events than have previously been
observed at LEP or Run I at the Tevatron (1992--1996).
Now let's look at what we've done with this harvest.

\section{Properties of \Bs\ Mesons }
Knowledge of the basic properties of the \Bs\ meson --
how it is produced,
how massive it is,
how long it lives,
how it decays --
is the foundation upon which we build all subsequent studies using the
particle. Accurate measurements of these quantities are thus
essential for the tests of the Standard Model, and the searches for its
extensions, described later. However, \Bs\ property determinations are
also useful in themselves as tests of our ability to use QCD,
whether through models or by lattice calculations.
Measurements made by CDF and \Dzero\ since the start of Run II have
substantially increased our knowledge in the full range of this area.

\subsection{Production Fraction}
Let's start with production. A major question here concerns whether
the fragmentation of quark-antiquark pairs into heavy flavor hadrons
is governed by universal functions independent of the type of
collisions producing the quarks. CDF has made preliminary measurements
of the relative fragmentation fractions of \Hadron{B}{} hadrons in \ppb\
collisions. 
Their preliminary measurement of the \Bs\ fraction relative to that of
\Hadron{B}{u} and \Hadron{B}{d}
uses \Bs \ra $\ell^- D_s^+ X$ decays in 0.36 \fbinv\ of data:
\begin{displaymath}
  \frac{f_s}{f_u + f_d} = 0.160 \pm 0.005 \; \mathrm{(stat)}
  \; ^{+0.011}_{-0.010} \; \mathrm{(syst)}
  \; ^{+0.057}_{-0.034} \; \mathrm{(BR)},
\end{displaymath}
where the last error reflects the uncertainty on the
$D_s^+ \ra \phi \pi^+$ branching ratio.
When corrected for $f_u$ and $f_d$, their measurement,
$f_s$ = (12.7 $\pm$ 3.8)\% is consistent with, but slightly higher
than the LEP average of (10.4 $\pm$ 0.9)\%\cite{hfag06},
indicating that no large differences in fragmentation between \epem\
and \ppb\ are likely.

\subsection{Mass and Lifetime}
Mass and lifetime are also important properties of the \Bs\ meson, 
and Run II results have
dramatically improved our knowledge of them. 
Before the start of Run II,
the \Bs\ mass world average of
5369.6 $\pm $ 2.4 MeV\cite{hfag03}
was dominated by 32 \BsJpsiPhi\ candidates reconstructed by CDF.
Using 0.22 \fbinv\ of data from Run II, 
corresponding to 185 candidate decays,
CDF has improved this measurement by almost a factor of three to 
5366.01 $\pm $ 0.73 $\pm$ 0.33 MeV\cite{cdfmass}.
On the lifetime side,
both CDF and \Dzero\ have contributed to a factor of two improvement
in the accuracy of the mean \Bs\ lifetime measurement 
since the start of Run II:
from $\tau (\Bs )$ = 1.461 $\pm$ 0.057 ps in 2003\cite{hfag03}
to $\tau (\Bs )$ = 1.451 $^{+0.029}_{-0.028}$ ps in 2006\cite{hfag06}.
Although each experiment has measured the \Bs\ lifetime in several
different modes, 
which contain different mixtures of \CP -even and \CP -odd states
(as discussed later in this article),
the most precise measurement,
$\tau (\Bs )$ = 1.398 $\pm$ 0.044 $^{+0.029}_{-0.028}$ ps,
currently comes from \Dzero 's analysis of
\Bs \ra \Ds $\mu X$ decays in 0.40 \fbinv\ of data\cite{d0life}.

Using the new average \Bs\ lifetime, we find agreement with
predictions of the lifetime ratio $\tau (\Bs) /\tau (B_d )$,
calculated using heavy quark effective theory, at the
2.3-sigma level:
\begin{center}
\begin{tabular}{ll}
  Experiment & 0.950 $\pm$ 0.019\cite{hfag06} \\
  NLO Theory & 1.00 $\pm$ 0.01\cite{tarantino}.
\end{tabular}
\end{center}
Note especially that the accuracy of the experimental measurements is
now approaching that of the theoretical predictions for this ratio.

\subsection{Hadronic Branching Ratios}
Making use of their 2-track trigger, CDF has been able to accumulate
an unprecedented sample of fully hadronic \Bs\ decays. This allows
them to make measurements, often for the first time ever, of various
rare mode branching ratios
as shown in tab. \ref{table:brs}.
These measurements provide valuable tests of QCD
models, particularly of the applicability of SU(3) quark
symmetries. With more statistics, some of them will also allow
sensitive tests of \CP\ violation:
the angle $\gamma / \phi_3$ of the CKM triangle from studies of
\Bs \ra $h^+ h^{\prime -}$,
and probes of \CP -even vs. \CP -odd contributions using
the $\psi(2S) \phi$ and $\phi \phi$ modes.

\begin{table}[ht]
\begin{center}
  \caption{\it Hadronic \Bs\ branching ratio measurements compared to
           theoretical expectations.}
\label{table:brs}
\vskip 0.1 in
\begin{tabular}{|l|c|c|l|}
\hline
  {\bf Mode} & {\bf Lumi} & {\bf Signal} & {\bf Measurement}\\
  & & & {\bf [Prediction]}\\
\hline \hline
  $\frac{\BR (B_s \ra D_s^- \pi^+)}{\BR (B^0 \ra D^- \pi^+)}$
  & 0.355 \fbinv & 494 $\pm$ 28 & 1.13 $\pm$ 0.08 $\pm$ 0.23\cite{cdfrdpi}\\
  & & & [1.05 $\pm$ 0.24]\cite{thrdpi} \\
\hline
  $\frac{\BR (B_s \ra D_s^- \pi^+ \pi^+ \pi^-)}
        {\BR (B^0 \ra D^- \pi^+ \pi^+ \pi^-)}$
  & 0.355 \fbinv & 309 $\pm$ 26 & 1.05 $\pm$ 0.10 $\pm$ 0.22\cite{cdfrdpi}\\
\hline
  \BR ($B_s \ra K^+ K^-$) ($\times 10^6$)
  & 1.0 \fbinv & 1307 $\pm$ 64 & 24.4 $\pm$ 1.4 $\pm$ 4.6 (prelim)\\
  & & & [20 $\pm$ 9]\cite{th1kk} \\
  & & & [35 $^{+73}_{-20}$]\cite{th2kk} \\
\hline
  \BR ($B_s \ra K^- \pi^+$) ($\times 10^6$)
  & 1.0 \fbinv & 230 $\pm$ 38 & 5.00 $\pm$ 0.75 $\pm$ 1.00 (prelim)\\
  & & & [4.9]\cite{thkpi} \\
\hline
  \BR ($B_s \ra \pi^+ \pi^-$) ($\times 10^6$)
  & 1.0 \fbinv & 26 $\pm$ 21 & $<$1.36 at 90\% CL (prelim)\\
\hline
  $\frac{\BR (B_s \ra \psi(2S) \phi)}{\BR (B_s \ra J/\psi \phi)}$
  & 0.36 \fbinv & 32.5 $\pm$ 6.5 
  & 0.52 $\pm$ 0.13 $\pm$ 0.07\cite{cdfpsi2s}\\
  & & & [0.54 $\pm$ 0.06]\cite{thpsi2s} \\
\hline
  \BR ($B_s \ra \phi \phi$) ($\times 10^6$)
  & 0.18 \fbinv & 7.3 $\pm$ 2.9 
  & 14 $^{+6}_{-5}$ $\pm$ 6\cite{cdfphiphi} \\
  & & & [10 -- 37]\cite{thphiphi}\\
\hline
\end{tabular}
\end{center}
\end{table}

Measurements of orbitally excited \Bs\ mesons have also been made by
CDF and \Dzero . However, these are discussed in another contribution
to these proceedings\cite{kuhr}.

\section{Flavor Changing Neutral Current Decays}
\begin{figure}[t]
\begin{center}
\begin{minipage}[t]{0.66\textwidth}
  \begin{center}
    \mygraphics{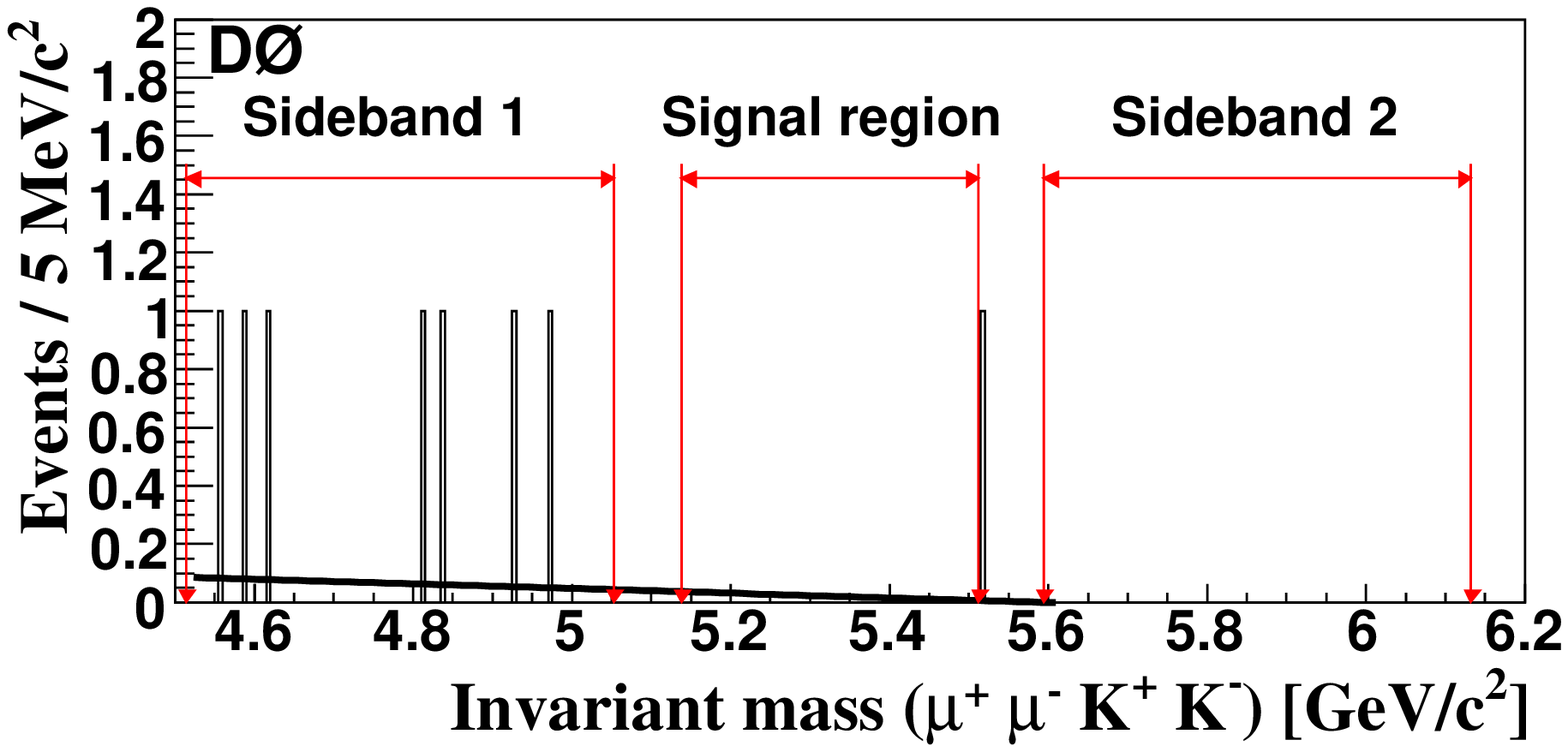}
  \end{center}
\end{minipage}
\hfill
\begin{minipage}[t]{0.33\textwidth}
  \begin{center}
    \mygraphics{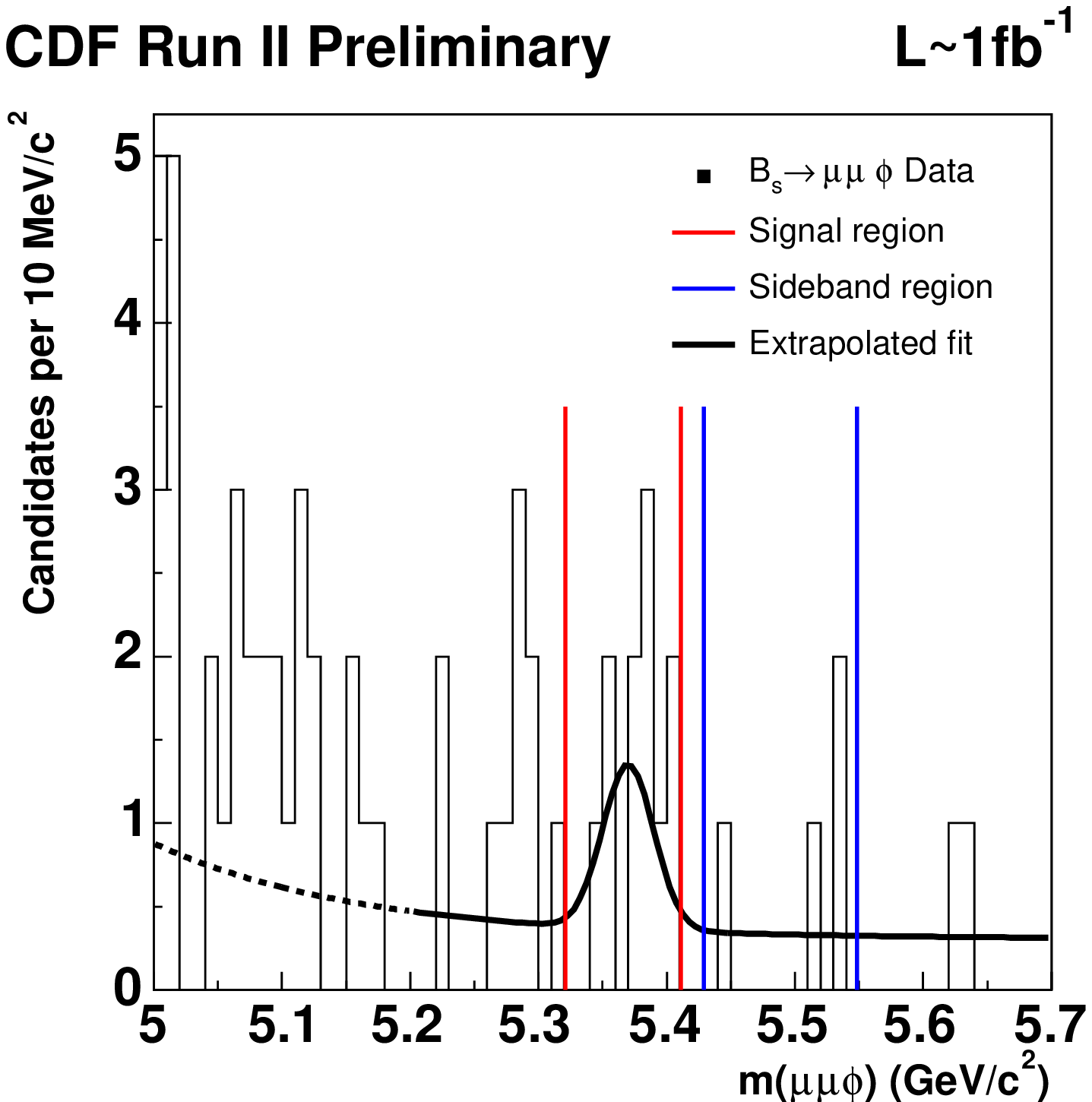}
  \end{center}
\end{minipage}
  \caption{\it The $\mu^+ \mu^- \phi$ invariant mass distribution observed
           by \Dzero\ (left) and CDF (right).}
\label{fig:mumuphi}
\end{center}
\end{figure}
Decays of \Hadron{B}{} hadrons governed by flavor changing neutral
currents provide sensitive probes for new physics
because these decays are highly suppressed in the Standard Model,
proceeding through loop diagrams.  In many models of physics beyond
the Standard Model, however, these types of decays can be enhanced in
some regions of model parameter space. For example, \BR (\Bsmumu ) is
proportional to $\tan^6 \beta$ \footnote{$\beta$ is the ratio of the
  vacuum expectation value of the two Higgs doublets.}
in the minimal supersymmetric standard model\cite{mssmmumu}.

CDF and \Dzero\ have searched for flavor changing neutral currents
in decays of \Bs\ mesons to $\mu^+ \mu^-$ and in
the decay \Bsmumuphi . Results are shown in
tab. \ref{table:fcnc}. The new limits on \BR (\Bsmumu ) 
represent more than an order of magnitude improvement over those
available before Run II data was analyzed\cite{hfag03}
and are now only a factor of 30 from the Standard Model prediction.
In the decay \Bsmumuphi , studied for the first
time with Run II data, CDF sees a 2.3-sigma excess of events,
as shown in fig. \ref{fig:mumuphi} -- so
observation of this mode could be just around the corner!

\begin{table}[ht]
\begin{center}
  \caption{\it Experimental limits and standard model predictions for
           flavor changing neutral current \Hadron{B}{} meson decays.}
\label{table:fcnc}
\vskip 0.1 in
\begin{minipage}{\textwidth}
\begin{tabular}{|l|c|c|c|c|l|}
\hline
  {\bf Mode} & {\bf Exp.} & {\bf Lumi} 
  & {\bf Evts} & {\bf Bgrd} & {\bf 95\% CL Limit}\\
\hline \hline
  \Bsmumu & \Dzero & 0.30 \fbinv & 4 & 4.3 $\pm$ 1.2 &
  $<$\scinot{4.0}{-7} (prelim)\footnote{Shortly after the end 
    of this conference \Dzero\ announced a new,
    preliminary \Bsmumu\ limit of \scinot{9.3}{-8} 
    at the 95\% CL, using 2 \fbinv\ of data.}\\
  & CDF & 0.78 \fbinv & 1 & 1.27 $\pm$ 0.37 & $<$\scinot{1.0}{-7} (prelim)\\
  & Pred. & & & & \scinot{(3.42$\pm$0.54)}{-9}\cite{thmumu}\\
\hline
  \Bsmumuphi & \Dzero & 0.45 \fbinv 
    & 0 & 1.6 $\pm$ 0.4 & $<$\scinot{4.1}{-6}\cite{d0mumuphi}\\
  & CDF & 0.92 \fbinv & 11 & 3.5 $\pm$ 1.5 & $<$\scinot{2.4}{-6} (prelim)\\
  & Pred. & & & & \scinot{(1.6$\pm$0.5)}{-6}\cite{thmumuphi}\\
\hline
\end{tabular}
\end{minipage}
\end{center}
\end{table}

\section{Mixing and \CP\ Violation}
The phenomenon of mixing between neutral mesons and anti-mesons
provides a very sensitive probe of the mechanism of electro-weak
symmetry breaking. This sensitivity is due to the fact that flavor
structure in the Standard Model,
in particular the difference between quark weak and mass eigenstates,
arises through Yukawa couplings of fermions to the Higgs
boson\cite{tevbwg}. Thus, time evolution in the neutral \Hadron{B}{}
systems is governed by the Schr\"{o}dinger equation:
\begin{equation}
  i \frac{d}{dt}
  \left( \begin{array}{c}
    | B(t) \rangle \\
    | \bar{B}(t) \rangle
  \end{array} \right)
  =
  \left( \begin{array}{cc}
    M - \frac{i\Gamma}{2} & M_{12} - \frac{i \Gamma_{12}}{2} \\
    M^*_{12} - \frac{i \Gamma^*_{12}}{2} & M - \frac{i\Gamma}{2}
  \end{array} \right)
  \left( \begin{array}{c}
    | B(t) \rangle \\
    | \bar{B}(t) \rangle
  \end{array} \right)
\end{equation}
and the eigenstates of the mass matrix, $B_L, B_H$, are different than
the weak eigenstates, $B, \bar{B}$, which oscillate between each
other.
These oscillations can be described by three parameters:
$|M_{12}|$, $|\Gamma_{12}|$, 
and the \CP -violating phase, $\phi = \mathrm{arg}(-M_{12}/\Gamma_{12})$,
which are related to physical observables:
\begin{equation}
\begin{array}{lclcl}
  \Delta m & = & M_H - M_L & \sim & 2 |M_{12}| \\
  \Delta \Gamma & = & \Gamma_L - \Gamma_H & = 
  & \Delta \Gamma_{CP} \cos \phi \\
  \Delta \Gamma_{CP} & = & \Gamma_{CP-even} - \Gamma_{CP-odd}
   & \sim & 2 |\Gamma_{12}| \\
\end{array}
\end{equation}

In the \Bs\ system, a measurement of the mass difference, \dms ,
which also gives the frequency of oscillations between particle and
anti-particle states, allows the determination of the CKM matrix element
\Vckm{ts}. Although important as parameters of the Standard Model,
\dms\ and \Vckm{ts} are most useful in constraining new physics when
used in conjunction the \Hadron{B}{d} oscillation frequency via:
\begin{equation}
  \frac{\dmd}{\dms} = \frac{M(\Hadron{B}{d})}{M(\Bs )} \xi
  \left| \frac{\Vckm{td}}{\Vckm{ts}} \right| ^2
  \hspace{0.5in}
  \xi = \frac{f^2_{B_d} B_{B_d}}{f^2_{B_s} B_{B_s}}
\end{equation}
The ratio of \dmd\ to \dms\ gives a more precise determination of
\Vckm{td}, related to one of the sides of the unitarity
triangle\cite{tevbwg}, because the theoretical uncertainty on $\xi$
is much less than that on the individual lattice calculations of the
\Hadron{B}{} meson decay constants ($f$) and bag parameters
($B$)\cite{xilattice}. 

The other \Bs\ mixing observables are also important in searches for
new physics. The ratio \dGs /\dms\ is a function of QCD parameters
only, and thus provides a measurement in this system that is
independent of new physics. The \CP -violating phase, \phis , however,
is expected to be tiny in the Standard Model, 
$\sim$0.25$^o$\cite{thdgamma}. 
Theories of physics
beyond the Standard Model, however, often include other sources of 
\CP -violation than the single Standard Model phase, and can thus
yield large predictions for \phis .

\subsection{\Bs\ Oscillations}
\begin{figure}[t]
\begin{center}
\begin{minipage}[b]{0.50\textwidth}
  \begin{center}
    \mygraphics{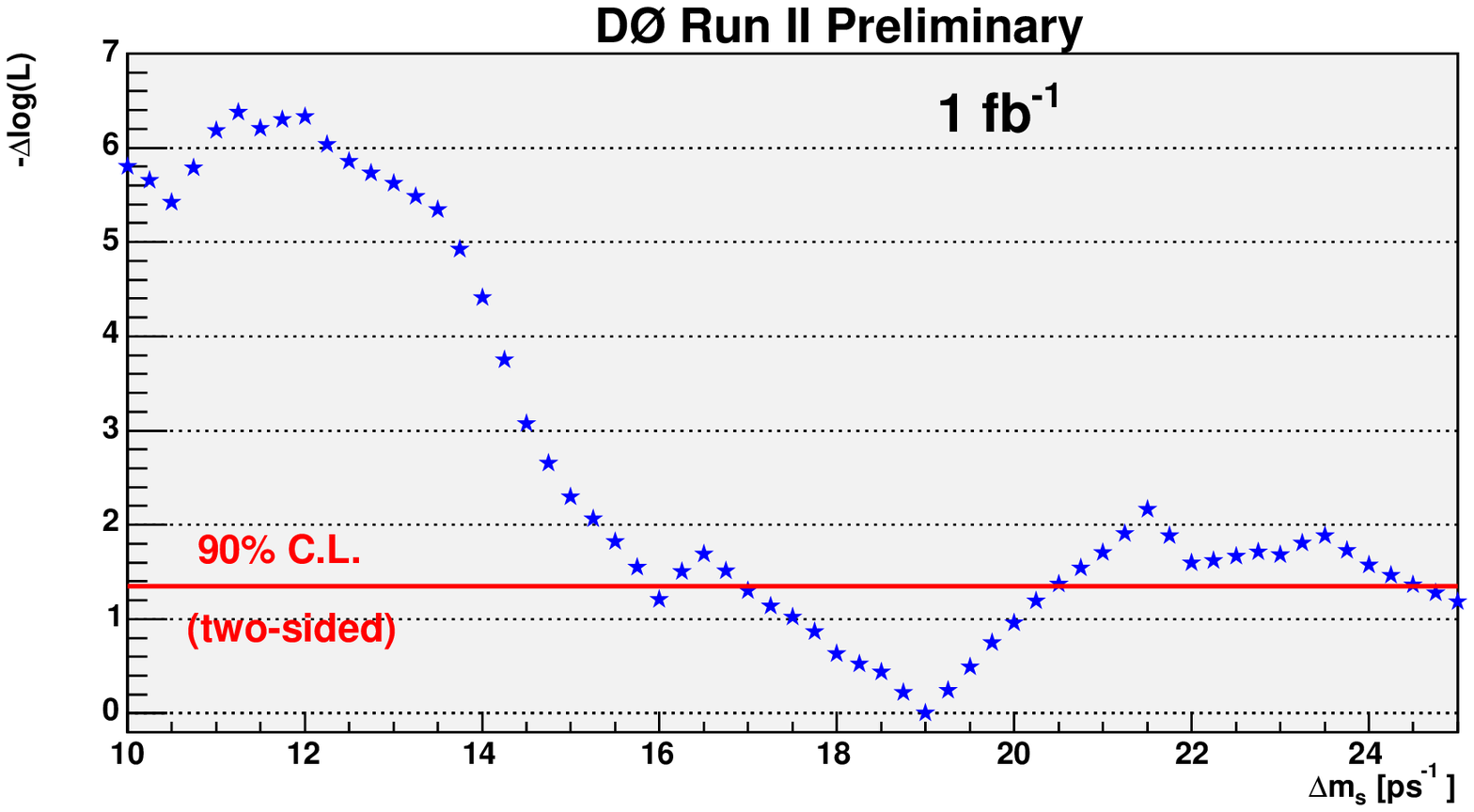}
    \vspace*{0.01in}
  \end{center}
\end{minipage}
\hfill
\begin{minipage}[b]{0.45\textwidth}
  \begin{center}
    \mygraphics{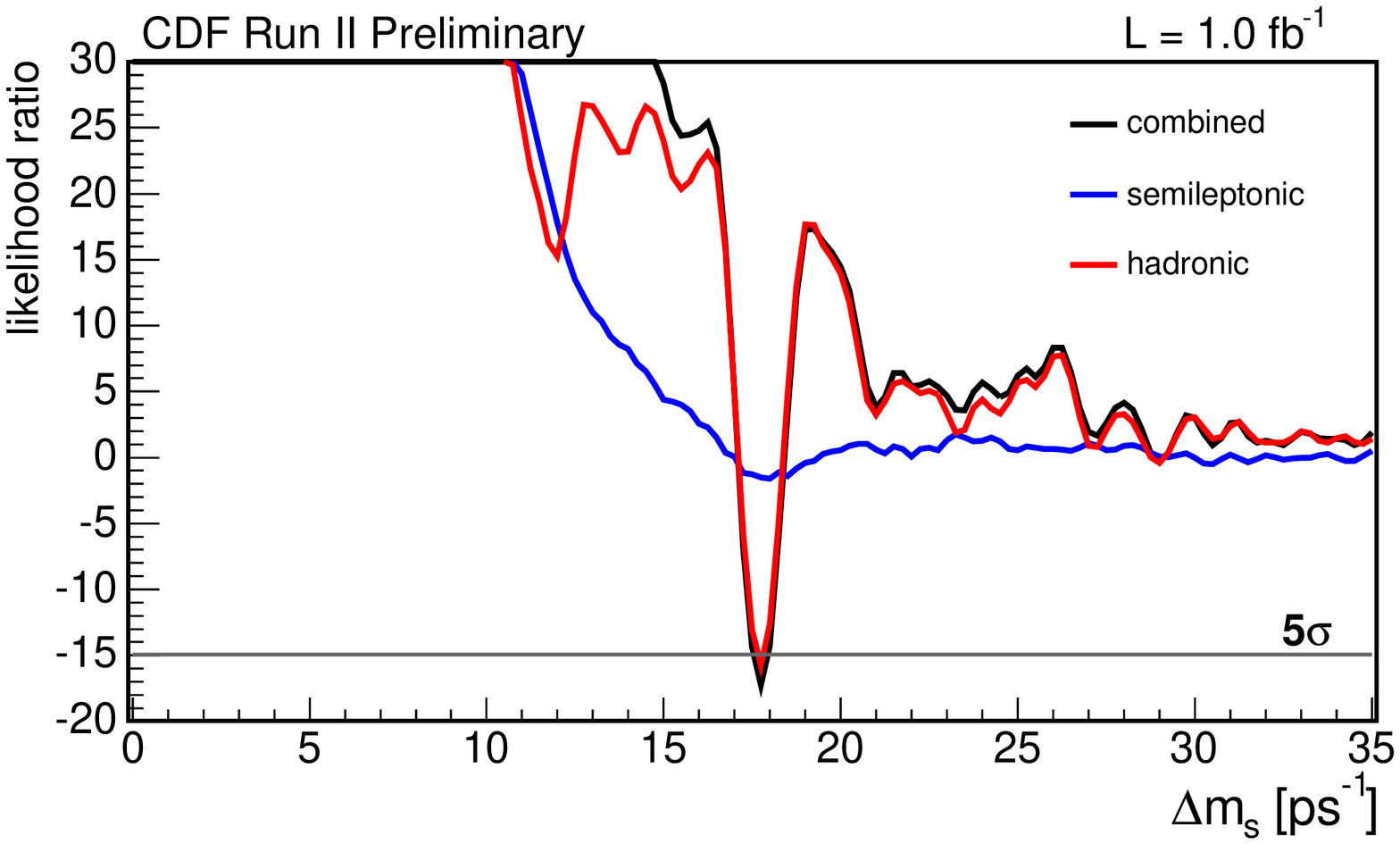}
  \end{center}
\end{minipage}
  \caption{\it Likelihood scans of the \Dzero\ (left) and CDF (right)
           combined oscillation samples vs. \dms .}
\label{fig:dms}
\end{center}
\end{figure}
Spring 2006 was a watershed period in the study for \Bs\
oscillations. After nearly two decades of searching for a \Bs\
oscillations signal by many experimental groups,
\Dzero\ was able to set the
first two-sided bound on the parameter \dms\ by a single 
experiment\cite{d0dms}. CDF followed quickly thereafter with
three-sigma evidence for \Bs\ oscillations\cite{cdfdmsev}.

The current status of \Bs\ mixing measurements is shown in
fig. \ref{fig:dms}
where a preliminary combination of \Dzero\ results is presented as
well as the final CDF observation\cite{cdfdmsmeas},
which is now significantly more than a 5-sigma effect.
Numerically, the results are:
\begin{center}
\begin{tabular}{lcl}
  \Dzero\ (1 \fbinv ) 
  & 17 $<$ \dms\ $<$ 21 \unitexp{ps}{-1} & (90\% CL) \\
  CDF (1 \fbinv )
  & 17.77 $\pm$ 0.10 $\pm$ 0.07 \unitexp{ps}{-1}.
\end{tabular}
\end{center}

Some details of the analyses are presented in tab. \ref{table:dms},
including the \Bs\ decay modes used; sample sizes; 
the quality of the
flavor taggers used to distinguish events where a \Bs\ oscillated to a
\Bsbar\ before decaying and vice-versa\footnote{
  This quality is defined as \effDsq ,
  the efficiency of the tagger times the
  dilution squared (where the dilution, $D = 1-2P_{mis-tag}$, 
  with $P_{mis-tag}$ being the probability
  to incorrectly tag the event).
  It is measured using information about the {\it other} \Hadron{B}{}
  hadron in the event (OST) or using particles associated with the
  \Bs\ meson (SST)
};
and the sensitivity of the analysis, defined as the expected limit in
the absence of any signal.
For reference, information about the previously most sensitive single
channel -- ALEPH's fully hadronic signal\cite{hfag03} -- is also
included.

CDF uses their measurement of \dms\ to derive a value for
the ratio of CKM matrix elements:
\begin{displaymath}
  \left| \frac{\Vckm{td}}{\Vckm{ts}} \right| =
  0.2060 \; \pm \; 0.0007 \mathrm{(exp)} \;
  ^{+0.0081}_{-0.0060} \mathrm{(theory)}.
\end{displaymath}
The accuracy on this quantity is now completely dominated by the
uncertainty of the ratio of \Bs\ and \Hadron{B}{d}
decay constants and bag parameters, $\xi$,
from lattice calculations\cite{xilattice}.

BaBar and Belle have also recently measured
$|\Vckm{td}/\Vckm{ts}|$ using \Hadron{B}{d} decays to $\rho \gamma$ and 
$K^* \gamma$\cite{nishida}. The average of their results,
0.200$\pm$0.016(exp)$^{+0.016}_{-0.015}$(theory), is consistent with
the CDF measurement but of significantly lower sensitivity.
Interestingly, however, the B-factory measurement is still dominated
by experimental uncertainties, while the theoretical error on its
value is only slightly larger than that on CDF's measurement. The
addition of enough data to bring the experimental uncertainty on the
B-factory measurements below that from theory will make this
measurement competitive with, and complimentary to, the matrix element
ratio measurements from the Tevatron.

\begin{table}[ht]
\begin{center}
  \caption{\it Details of the \Dzero\ and CDF \Bs\ oscillation
           analyses. Also shown are comparable numbers for the
           previously most sensitive analysis from ALEPH.}
\label{table:dms}
\vskip 0.1 in
\begin{tabular}{|lcccc|}
\hline
  {\bf Mode} & {\bf Sample} 
  & \multicolumn{2}{c}{\bf Average \effDsq} & {\bf Sensitivity}\\
  & & {\bf OST} & {\bf SST} &\\
\hline \hline
  ALEPH & 28.5 & \multicolumn{2}{c}{27\%} & 13.6 \unitexp{ps}{-1}\\
  \multicolumn{5}{|l|}{fully hadronic}\\
\hline
  \Dzero\ Semileptonic & 43,000 & 2.48\% & & 16.5 \unitexp{ps}{-1}\\
  \multicolumn{5}{|l|}
	      {$\ell$\Ds ; \Ds \ra $\phi\pi^- , K^{*0}K^- ,K^0_S K^-$}\\
\hline
  CDF Semileptonic & 61,500 & 1.8\% & 4.8\% & 19.3 \unitexp{ps}{-1}\\
  \multicolumn{5}{|l|}
	      {$\ell$\Ds ; \Ds \ra $\phi\pi^- , K^{*0}K^- ,3\pi^{\pm}$}\\
\hline
  CDF Hadronic & 8,700 & 1.8\% & 3.7\% & 30.7 \unitexp{ps}{-1}\\
  \multicolumn{5}{|l|}
	      {\Ds $\pi^+$, \Ds $3\pi^{\pm}$ ; 
	       \Ds \ra $\phi\pi^- , K^{*0}K^- ,3\pi^{\pm}$}\\
\hline
\end{tabular}
\end{center}
\end{table}

\subsection{\Bs\ Width Difference and \CP -violating Phase}
\begin{figure}[t]
\begin{center}
\begin{minipage}[b]{0.30\textwidth}
  \begin{center}
    \mygraphics{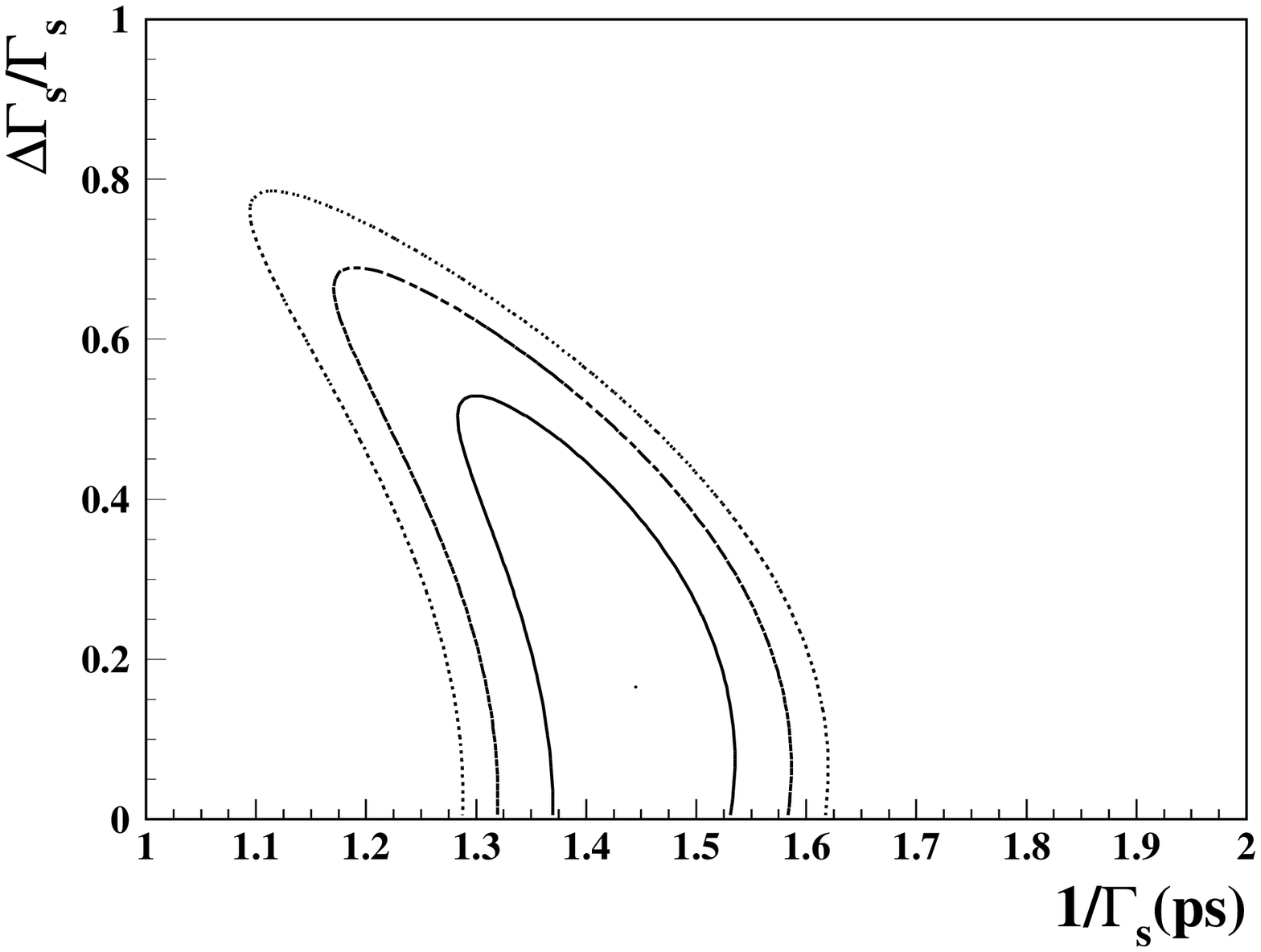}
    \vspace{0.15in}
  \end{center}
\end{minipage}
\begin{minipage}[b]{0.30\textwidth}
  \begin{center}
    \mygraphics{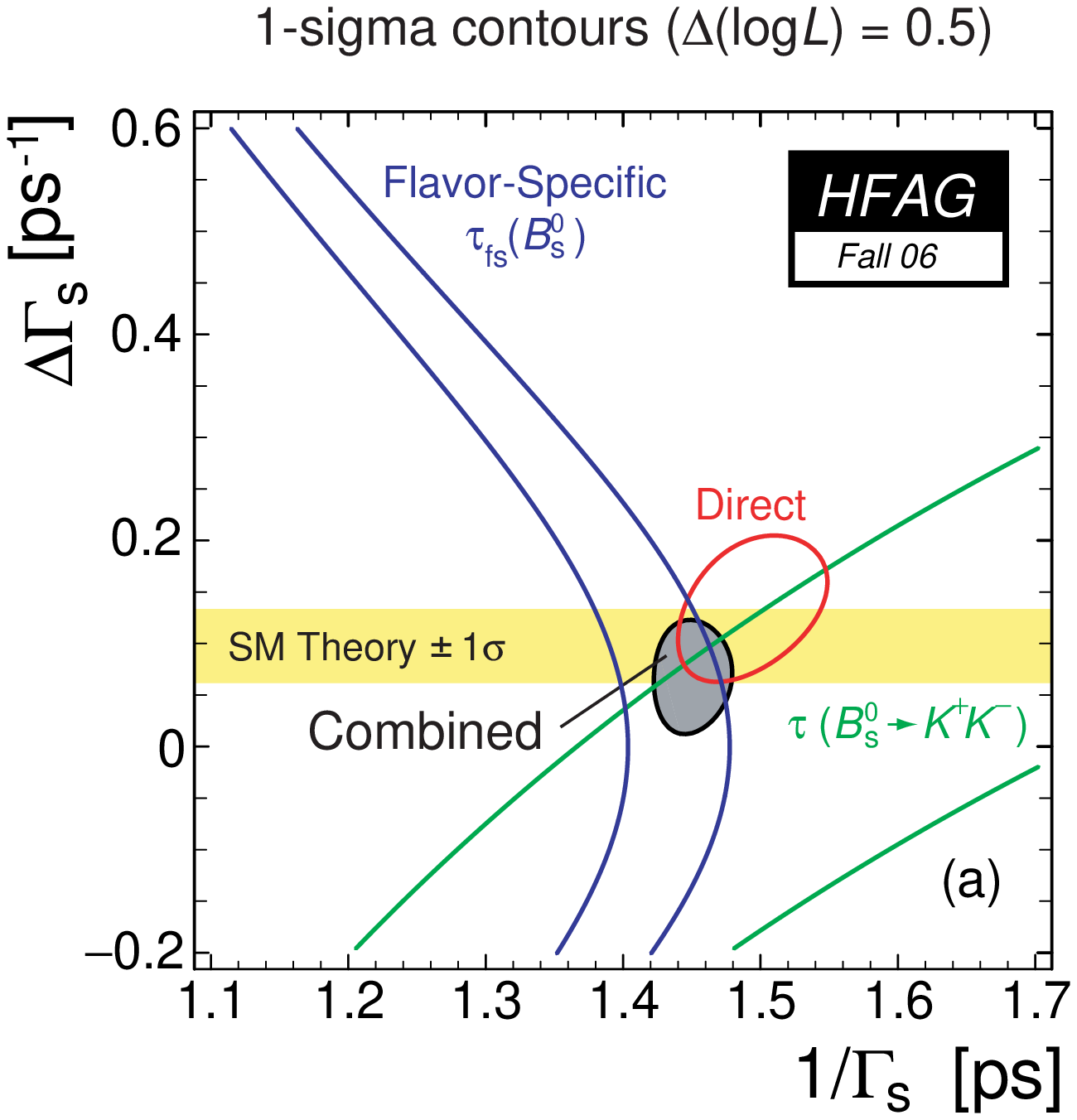}
  \end{center}
\end{minipage}
\begin{minipage}[b]{0.30\textwidth}
  \begin{center}
    \mygraphics{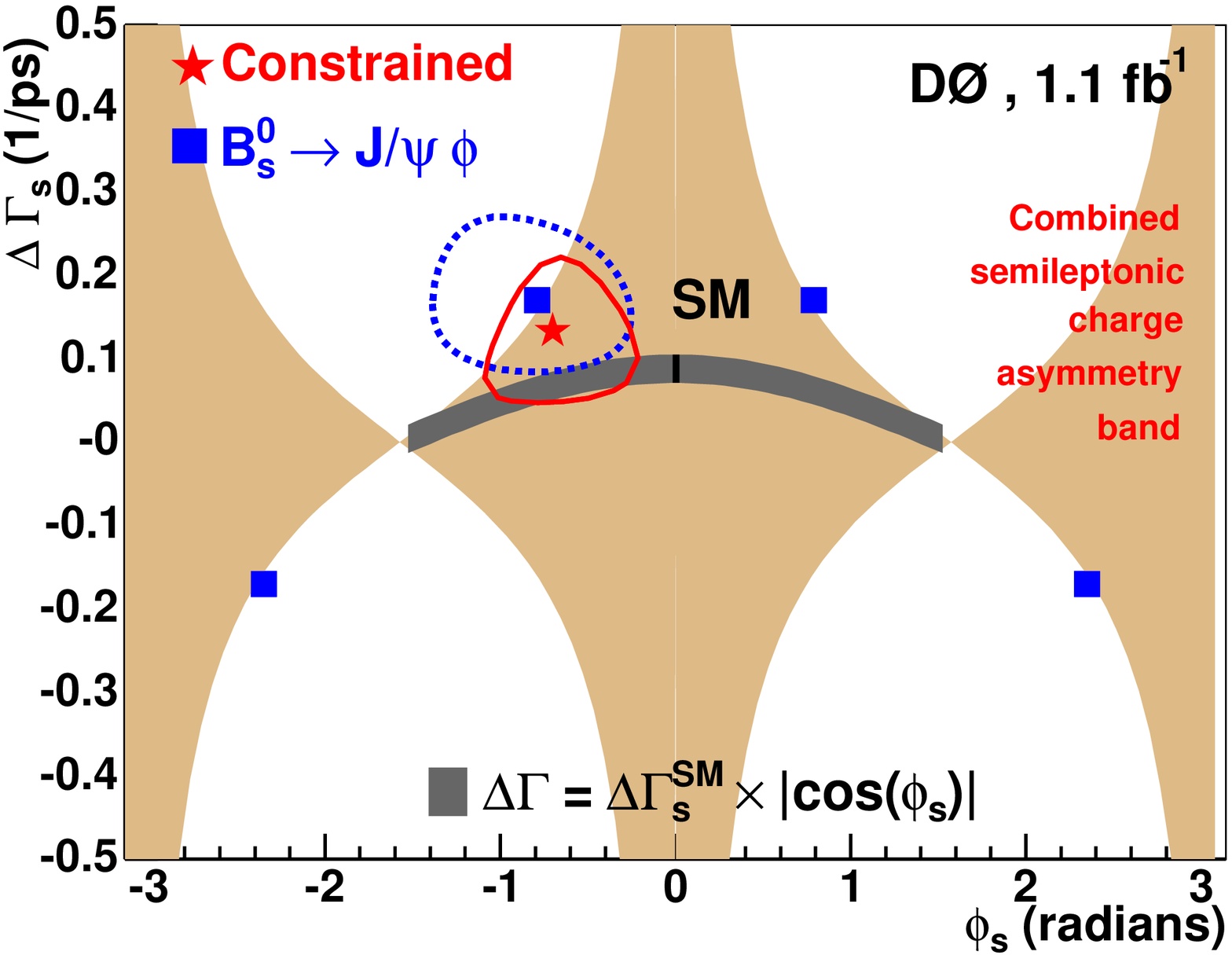}
    \vspace{0.01in}
  \end{center}
\end{minipage}
  \caption{\it Plots of the world average values of \dGs\
           vs. 1/$\Gamma_s$ from winter 2003\cite{hfag03} (left) and
	   the end of 2006\cite{hfag06} (middle).
	   Also show is the combination\cite{d0dgcomb} 
	   of \Dzero\ measurements of
           \dGs\ vs \phis\ (right) .}
\label{fig:dgamma}
\end{center}
\end{figure}
Not content with measuring only the mass difference part of \Bs\
mixing, intrepid analysts also made major progress on the other
parameters, \dGs\ and \phis , in the last year. These quantities can
be accessed by combining information from several sources:
\begin{enumerate}
  \item \BsJpsiPhi : the time evolution, mass and angular distributions in
        \BsJpsiPhi\ decays;
  \item \taufs : lifetimes of flavor-specific \Bs\ decays, for example
        semi-leptonic decays, which contain an equal mixture of \CP
        -even and \CP -odd components;
  \item \Aslmu : charge asymmetries in like-sign dimuon production;
  \item \Asls : charge asymmetries in \Bs \ra $\mu^{\pm} D_s^{\mp}$ decays;
  \item \taueven , \BReven : 
        lifetimes and branching ratios of \CP -specific \Bs\ decays,
        such as \Bs \ra $K^+ K^-$ and \BsDsDs .
\end{enumerate}
In the past, analyses centered on the extraction of \dGs\ using the
first two methods described above. The state-of-the-art in early 2003
can be seen in the left-most plot of fig. \ref{fig:dgamma}, which
shows that no statistically significant extraction of the value of
\dGs\ could be made before Run II results were available.

This has changed dramatically in the past year, with a flurry of new
results from \Dzero\ and CDF, which are summarized in
tab. \ref{table:dgamma}. Combining these results 
(with the exception
of the \Dzero\ and CDF \BsDsDs\ measurements where the assumption of
\CP -even state dominance is unproven) 
yields averages summarized in
the center plot of fig. \ref{fig:dgamma}\cite{hfag06}. Progress is
clearly substantial, with a new world-average value of
\begin{displaymath}
  \dGs = 0.071 ^{+0.053}_{-0.057} \; \unitexp{ps}{-1}
  \hspace{0.5in}
  (-0.04 < \dGs < +0.17) \; \unitexp{ps}{-1} \;\; (95\% \; \mathrm{CL}).
\end{displaymath}
This measurement, favors a positive, non-zero value for \dGs ,
and is in agreement with the Standard Model expectation of
0.088$\pm$0.017 \unitexp{ps}{-1}\cite{thdgamma}.

Their large \Bs\ data samples and multiple handles on mixing also allow
\Dzero\ to perform a combination of their results\cite{d0dgcomb},
shown in the right-hand plot of fig. \ref{fig:dgamma},
that is sensitive to \phis . This combination
results in a value of \dGs = 0.13$\pm$0.09 \unitexp{ps}{-1}, which is
consistent with the world average;
and finds:
\begin{displaymath}
  \phis = -0.70 ^{+0.47}_{-0.39}
\end{displaymath}
which is nearly 2-sigma away from
\scinot{(4.2$\pm$1.4)}{-3}\cite{thdgamma},
the Standard Model prediction.

\begin{table}[htp]
\begin{center}
  \caption{\it A summary of recent analyses sensitive to 
           \dGs\ and \phis .}
\label{table:dgamma}
\vskip 0.1 in
\begin{tabular}{|l|l|l|l|}
\hline
  {\bf Tech.} & {\bf Exp.} & {\bf Observables} & {\bf Sens. to \dGs , \phis }\\
  & {\bf Lumi} & {\bf Measurement} & \\
  & {\bf Signal} & & \\
\hline \hline
  \BsJpsiPhi & CDF\cite{cdfpsiphi} 
  & $M(J/\psi \phi)$, proper time,
  & fit for: \dGs , \taufs \\
  & 0.355 \fbinv & 3 decay angles & helicity amplitudes\\ 
  & 203$\pm$15 & \dGs\ = 0.47$^{+0.19}_{-0.24}\pm$0.01 \unitexp{ps}{-1} & \\
\hline
  \BsJpsiPhi & \Dzero \cite{d0psiphi} 
  & $M(J/\psi \phi)$, proper time,
  & fit for: \dGs , \phis , \taufs \\
  & 1.1 \fbinv & 3 decay angles & helicity amplitudes,\\ 
  & 1039$\pm$45 & \dGs\ = 0.17$\pm$0.09 \unitexp{ps}{-1} & strong phases\\
  & & \phis\ = $-$0.79$\pm$0.56 & \\
\hline
  \taufs & W.A.\cite{hfag06}
  & \taufs\ = 1.440$\pm$0.036 ps & \taufs \\
  & &
  & \(
      = \dfrac{1}{\Gamma_s} 
      \left[ \dfrac{1 + (\dGs /2\Gamma_s)^2}{1 - (\dGs /2\Gamma_s)^2}
      \right]
    \) \\
\hline
  \Aslmu & \Dzero \cite{d0aslmu}
  & \(
    \dfrac{N(b\bar{b} \ra \mu^+\mu^+)-N(b\bar{b} \ra \mu^-\mu^-)} 
          {N(b\bar{b} \ra \mu^+\mu^+)-N(b\bar{b} \ra \mu^-\mu^-)} 
    \)
  & \( \Aslmu = A_{SL}^{d} + \dfrac{f_s Z_s}{f_d Z_d}\Asls \)\\
  & 1.0 \fbinv 
  & \Aslmu = (-0.92$\pm$0.44$\pm$0.32)\%
  & \(
      Z_q = \dfrac{1}{1 - (\Delta \Gamma_q /2\Gamma_q)^2} \)\\
  & & 
  & \(
    \; \; \; \; \; - \dfrac{1}{1 + (\Delta m_q /\Gamma_q)^2}
    \) \\
\hline
  \Asls & \Dzero \cite{d0asls}
  & \(
      \dfrac{N(\mu^+ D_s^-) -  N(\mu^- D_s^+)}
            {N(\mu^+ D_s^-) +  N(\mu^- D_s^+)}
    \)
  & \Asls \\
  & 1.3 \fbinv 
  & \Asls = (1.23$\pm$0.97$\pm$0.17)\% 
  & \( =
      \dfrac{1}{2} \dfrac{x_s^2 + y_s^2}{1+x_s^2}
      \dfrac{\dGs}{\dms} \tan \phis
    \) \\
  & 27,300$\pm$300 & & \\
\hline
  \taueven & CDF prelim
  & $\tau$(\Bs \ra $K^+ K^-$)
  & \taueven \\
  & 360 \fbinv & 1.53$\pm$0.18$\pm$0.02 ps 
  & \( 
      \sim \dfrac{1}{\Gamma_s}
      \left[ \dfrac{1}{1 + (\dGcp /2\Gamma_s)} \right]
    \) \\
  & 718$\pm$55 & & \\
\hline
  \BReven & CDF prelim
  & \BR(\Bs \ra $D_s^+ D_s^-$)
  & 2\BReven \\ 
  & 360 \fbinv & = (1.3$\pm$0.6)\% 
  & \( 
      \sim \dfrac{\dGcp}{\Gamma_s}
      \left[ \dfrac{1}{1 + (\dGcp /2\Gamma_s)} \right]
    \) \\
  & 718$\pm$55 & & \\
\hline
  \BReven & \Dzero \cite{d0dsds}
  & \BR(\BsDsDs ) & \\
  & 1.3 \fbinv & = (3.9$^{+1.9\;+1.6}_{-1.7\;-1.5}$)\% & \\
  & 718$\pm$55 & & \\
\hline
\end{tabular}
\end{center}
\end{table}

\subsection{Direct \CP - violation}
As a final note to this section on \CP\ measurements using
the \Bs , CDF has taken the first steps toward measuring direct 
\CP -violation in the \Bs\ system 
using their preliminary \Bs \ra $K^- \pi^+$
measurement, discussed previously. In addition to measuring the
branching ratio for this mode, they also determine its \CP\
asymmetry:
\begin{displaymath}
  A_{CP} \equiv
  \frac{|A(B^0_s \ra K^-\pi^+)|^2 - |A(\bar{B}^0_s \ra K^+\pi^-)|^2}
  {|A(B^0_s \ra K^-\pi^+)|^2 + |A(\bar{B}^0_s \ra K^+\pi^-)|^2}
  = 0.39 \; \pm \; 0.15 \; \pm \; 0.08,
\end{displaymath}
which differs from zero by 2.5-sigma and is in good agreement with the
Standard Model expectation of $\sim$0.37\cite{thAcp}.

\section{Future Prospects and Conclusions}
Looking back on the last few years, 
we can take pride in the good use to which the \Bs\ meson has been put
at the Tevatron in Run II. We have observed many decay modes of this
particle for the first time and are zeroing in on an observation of
the flavor changing neutral current in \Bsmumuphi\ decays,
while being only a factor of 30 away from the Standard Model
prediction for \Bsmumu .
We have also made remarkable progress in our understanding of \Bs\
mixing, with a first observation of its oscillation frequency after
more than a decade of searching;
and new sensitivity to the \CP -violating phase in this system.
However, we certainly do not plan to rest on our laurels.

The Tevatron is operating extremely well, delivering luminosity at a
pace where we can expect a total Run II data sample of up to 8
\fbinv\ per experiment. In addition, both experiments are upgrading
their capabilities, with \Dzero 's new Layer-0 silicon detector of
particular importance to the B-physics program. Although the larger
instantaneous luminosities seen by CDF and \Dzero\ will force the
imposition of more restrictive triggers,
significantly larger \Bs\ data samples should be available in the next
1--2 years. As many of the measurements presented here remain
statistics limited
(a notable expectation is the extraction of \Vckm{td} from \dmd\ and
\dms )
this added data should allow a vibrant continuing program of
measurements of rare decay modes and \CP -violation in the \Bs\
system. 

You have, most certainly, not heard the last of the \Bs\ meson at the
Tevatron!

\section*{Acknowledgments}
I would like to thank 
the \Dzero\ and CDF B-physics working group convenors,
Brendan Casey, Cheng-Ju Lin, Manfred Paulini, and Andrzej Zieminski,
for their help in preparing this presentation. 
My warmest gratitude also goes to the conference organizers for
preparing such a fascinating program
(not to mention the excellent food, drink, and skiing)!


\end{document}